Title: Magnetic properties of collinear structures of compound La1-x CaxMnO3

Authors: B. V. Karpenko (1), L. D. Falkovskaya (1), and A. V. Kuznetsov (2)

( (1) Institute of Physics of Metals, Ural Branch of Russian Academy of Sciences, Ekaterinburg, Russia, (2) Ural State University, Ekaterinburg, Russia )

With use of homogeneous model for magnetic sub-lattice of compound La1-x CaxMnO3 ($0 \le x \le 1$) allowing for Heisenberg superexchange and double exchange between nearest neighbors the temperature dependence of spontaneous magnetization on a separate site is investigated in molecular field approximation. The difference from the usual magnets with one type of magnetic ions is in the fact that system of equations appears with two unknowns , being the mean spin values of ions $Mn3+$ and $Mn4+$, correspondingly. Near ordering temperature this system can be solved analytically and at the same time the expression for the ordering temperature can be obtained. To find the temperature dependence of magnetization in the whole temperature range we used the exact expression for double exchange operator containing spin operators under the square root. The numerical solutions were found in this case.

# I. INTRODUCTION

Paper [1] contained theoretical investigation of a question about appearing different magnetic structures in the row of compounds $La_{1-x}Ca_xMnO_3$, $0 \le x \le 1$ (the structure type $GdFeO_3$) at absolute zero temperature. The homogeneous model of magnetic manganese sub-lattice was used that is it was considered that each site of this sub-lattice is with probability $x$ occupied by ion $Mn^{4+}$ and has spin 3/2 and with probability $(1-x)$ – by ion $Mn^{3+}$ with spin 2. The problem was solved in the nearest neighbor approximation allowing for two interaction mechanisms -

Heisenberg superexchange and non-Heisenberg double exchange. Minimization of the ground state energy at T=0 with respect to the directing angles of magnetic sub-lattices resulted in the system of transcendental equations. The solutions of this system gave 11 types of magnetic configurations: two ferromagnetic, three collinear antiferromagnetic and six non-collinear antiferromagnetic. When the concentration of Ca ions x changes one spin configuration replaces another as the ground state. As a whole the sequence of configurations when $x$ changes from 0 till 1 can be brought in correspondence to those observed on the experiment.

The present work investigates magnetic properties of compound $La_{1-x}Ca_xMnO_3$ at temperatures not equal to zero in molecular field approximation. The consideration is carried out for four collinear magnetic configurations: ferromagnetic phase B and three antiferromagnetic phases A, C and G (in notations of paper[2]). The expressions for ordering temperatures of listed magnetic phases are found together with temperature dependences of spontaneous magnetization of a separate site near ordering temperatures and in the whole temperature range.

## II. HAMILTONIAN OF THE CRYSTAL

Like in paper [1], magnetic system is described by the model of homogeneous dynamic alloy of rigid spins $S_1 = 3/2\,(Mn^{4+})$ and $S_2 = 2\,(Mn^{3+})$. It is considered that $La_{1-x}Ca_xMnO_3$ in the whole range $0 \le x \le 1$ has the structure type $GdFeO_3$ (space group Pnma). The elementary cell of orthorhombic lattice is presented on Fig.1 where only manganese ions sites are presented. Numbers 1,2,3,4 enumerate four Bravais lattices. The base vectors of Bravais lattices (sub-lattices) are

$$\vec{\rho}_1 = 0, \vec{\rho}_2 = \frac{1}{2}(\vec{a} + \vec{b}), \vec{\rho}_3 = \frac{1}{2}\vec{c}, \vec{\rho}_4, = \frac{1}{2}(\vec{a} + \vec{b} + \vec{c}), \quad (1)$$

where $\vec{a}, \vec{b}$ and $\vec{c}$ are vectors of primitive translations for a simple orthorhombic lattice. Vectors of six nearest neighbors are given by the expressions

$$\vec{\Delta}_1 = \frac{1}{2}(\vec{a} + \vec{b}), \vec{\Delta}_2 = \frac{1}{2}(\vec{a} - \vec{b}), \vec{\Delta}_3 = -\frac{1}{2}(\vec{a} + \vec{b}), \vec{\Delta}_4 = \frac{1}{2}(-\vec{a} + \vec{b}), \vec{\Delta}_5 = \frac{1}{2}\vec{c}, \vec{\Delta}_6 = -\frac{1}{2}\vec{c}. \quad (2)$$

Let's denote the integral of superexchange interaction between ions $Mn^{3+}-Mn^{3+}$ in $(ab)$- plane as $I_1$ (the nearest neighbors $\vec{\Delta}_1,\vec{\Delta}_2,\vec{\Delta}_3,\vec{\Delta}_4$), while that along $c$-axis as $I_2$( the nearest neighbors $\vec{\Delta}_5$ and $\vec{\Delta}_6$). Similarly for interaction $Mn^{4+}-Mn^{4+}$ in $(ab)$-plane as $I_3$, while along $c$-axis as $I_4$. Interactions $Mn^{3+}-Mn^{4+}$ in $(ab)$-plane we shall denote as $I_5$, and along $c$-axis as $I_6$. The transfer integral between ions $Mn^{3+}$ and $Mn^{4+}$ in $(ab)$-plane let's be $B_1$, and along $c$ - $B_2$.

As a result the model spin Hamiltonian can be presented in the form

$$\hat{H}=\hat{H}_{1ex}+\hat{H}_{2ex}, \tag{3}$$

where

$$\hat{H}_{1ex}=-\sum_{\vec{m}}\sum_{k=1}^{4}\sum_{i=1}^{6}\sum_{n,l=1}^{2}I_{nl}(\vec{\Delta}_i)\vec{S}_n(\vec{m}+\vec{\rho}_k)\vec{S}_l(\vec{m}+\vec{\rho}_k+\vec{\Delta}_i), \tag{4}$$

$$\hat{H}_{2ex}=-\sum_{\vec{m}}\sum_{k=1}^{4}\sum_{i=1}^{6}\sum_{n,l=1}^{2}B_{nl}(\vec{\Delta}_i)\sqrt{1+\frac{2}{1+S(2S+3)}\vec{S}_n(\vec{m}+\vec{\rho}_k)\vec{S}_l(\vec{m}+\vec{\rho}_k+\vec{\Delta}_i)}\ . \tag{5}$$

Here $\hat{H}_{1ex}$ is the operator of superexchange interaction, $\hat{H}_{2ex}$ is the operator of double exchange. Summing over $\vec{m}$ in Eqs. (4-5) is led upon $N$ sites of sub-lattice (the whole number of sites is $4N$). Sum over index $k$ means summing over four sub-lattices. Index $i$ numbers the nearest neighbors. Indexes $n$ and $l$ distinguish ions $Mn^{3+}$ ($n,l=1$) and $Mn^{4+}$ ($n,l=2$). In this case $S_1=S+(1/2)$ (spin of ion $Mn^{3+}$) and $S_2=S$ (spin of ion $Mn^{4+}, S=3/2$). The following notations are introduced also in Eqs. (4-5):

$$I_{11}(\vec{\Delta}_i)=(1-x)^2I_1; i=1,2,3,4. \tag{6}$$

$$I_{11}(\vec{\Delta}_i)=(1-x)^2I_2; i=5,6. \tag{7}$$

$$I_{22}(\vec{\Delta}_i)=x^2I_3; i=1,2,3,4. \tag{8}$$

$$I_{22}(\vec{\Delta}_i)=x^2I_4; i=5,6. \tag{9}$$

$$I_{12}(\vec{\Delta}_i)=I_{21}(\vec{\Delta}_i)=x(1-x)I_5; i=1,2.3.4. \tag{10}$$

$$I_{12}(\vec{\Delta}_i)=I_{21}(\vec{\Delta}_i)=x(1-x)I_6; i=5,6. \tag{11}$$

$$B_{11}(\vec{\Delta}_i)=B_{22}(\vec{\Delta}_i)=0. \tag{12}$$

$$B_{12}(\vec{\Delta}_i) = B_{21}(\vec{\Delta}_i) = x(1-x)\sqrt{\frac{S+1}{2S+1}}B_1; i = 1,2,3,4. \tag{13}$$

$$B_{12}(\vec{\Delta}_i) = B_{21}(\vec{\Delta}_i) = x(1-x)\sqrt{\frac{S+1}{2S+1}}B_2; i = 5,6. \tag{14}$$

## III. TEMPERATURE DEPENDENCE OF MAGNETIZATION NEAR ORDERING TEMPERATURES FOR COLLINEAR MAGNETIC CONFIGURATIONS

It was shown in paper [1] that at certain values of ions Ca concentration $x$ the role of the ground state of the system is played by one of the following collinear magnetic configurations (in Wollan and Koehler [2] notations):

Antiferromagnetic phase A where each spin is surrounded by four parallel to it spins in (ab)-plane and by two antiparallel ones – along c-axis ($0 < x < 0.17$).

Ferromagnetic phase B (the range of existence is $0.2 < x < 0.5$).

Antiferromagnetic phase C where the environment consists of two parallel spins along c-axis and by four antiparallel ones – in (ab)-plane ($0.7 < x < 0.85$).

Antiferromagnetic phase G where all six nearest neighboring spins are antiparallel to the central one ($0.85 < x < 1$).

All collinear magnetic configurations are presented on Fig.2. Let's denote $\hat{S}_l(k)$ $(l = 1,2, k = 1,2,3,4)$ the spin operator $S_l$, belonging to the sub-lattice with number $k$. If one expands the square root in Hamiltonian $\hat{H}_{2ex}$ (5) up to the first degree of spin operators scalar product then in Weiss molecular field approximation we obtain

$$\langle S_l^z(k)\rangle = S_l B_{S_l}(Z_{lk}), \tag{15}$$

where

$$B_S(x) = \frac{2S+1}{2S} cth(\frac{2S+1}{2S}x) - \frac{1}{2S} cth(\frac{x}{2S}) \tag{16}$$

is the Brillouin function. Let's introduce the following notations for spins of ions $Mn^{3+}$ and $Mn^{4+}$, being in one of four magnetic sub-lattices:

$\langle S_1^z(1)\rangle = x_1$ $\langle S_1^z(2)\rangle = x_3$ $\langle S_1^z(3)\rangle = x_5$ $\langle S_1^z(4)\rangle = x_7$

$$\langle S_2^z(1)\rangle = x_2 \qquad \langle S_2^z(2)\rangle = x_4 \qquad \langle S_2^z(3)\rangle = x_6 \qquad \langle S_2^z(4)\rangle = x_8 . \tag{17}$$

Then we obtain the following system of equations for sub-lattices magnetizations

$$\begin{aligned}
x_1 &= S_1 B_{S_1}(Z_{11}) \\
x_2 &= S_2 B_{S_2}(Z_{21}) \\
x_3 &= S_1 B_{S_1}(Z_{12}) \\
x_4 &= S_2 B_{S_2}(Z_{22}) \\
x_5 &= S_1 B_{S_1}(Z_{13}) \\
x_6 &= S_2 B_{S_2}(Z_{23}) \\
x_7 &= S_1 B_{S_1}(Z_{14}) \\
x_8 &= S_2 B_{S_2}(Z_{24})
\end{aligned} \tag{18}$$

The following notations for molecular fields are accepted here:

$$\begin{aligned}
Z_{11} &= \frac{2S_1}{T}\{(1-x)^2[4I_1x_3 + 2I_2x_5] + x(1-x)[4\tilde{I}_5x_4 + 2\tilde{I}_6x_6]\} \\
Z_{21} &= \frac{2S_2}{T}\{x^2[4I_3x_4 + 2I_4x_6] + x(1-x)[4\tilde{I}_5x_3 + 2\tilde{I}_6x_5]\} \\
Z_{12} &= \frac{2S_1}{T}\{(1-x)^2[4I_1x_1 + 2I_2x_7] + x(1-x)[4\tilde{I}_5x_2 + 2\tilde{I}_6x_8]\} \\
Z_{22} &= \frac{2S_2}{T}\{x^2[4I_3x_2 + 2I_4x_8] + x(1-x)[4\tilde{I}_5x_1 + 2\tilde{I}_6x_7]\} \\
Z_{13} &= \frac{2S_1}{T}\{(1-x)^2[4I_1x_7 + 2I_2x_1] + x(1-x)[4\tilde{I}_5x_8 + 2\tilde{I}_6x_2]\} \\
Z_{23} &= \frac{2S_2}{T}\{x^2[4I_3x_8 + 2I_4x_2] + x(1-x)[4\tilde{I}_5x_7 + 2\tilde{I}_6x_1]\} \\
Z_{14} &= \frac{2S_1}{T}\{(1-x)^2[4I_1x_5 + 2I_2x_3] + x(1-x)[4\tilde{I}_5x_6 + 2\tilde{I}_6x_4]\} \\
Z_{24} &= \frac{2S_2}{T}\{x^2[4I_3x_6 + 2I_4x_4] + x(1-x)[4\tilde{I}_5x_5 + 2\tilde{I}_6x_3]\},
\end{aligned} \tag{19}$$

where

$$\tilde{I}_5 = I_5 + \frac{B_1}{1 + S_2(2S_2+3)}\sqrt{\frac{S_2+1}{2S_2+1}} \quad , \tag{20}$$

$$\tilde{I}_6 = I_6 + \frac{B_2}{1 + S_2(2S_2 + 3)}\sqrt{\frac{S_2 + 1}{2S_2 + 1}} \quad . \tag{21}$$

To find the temperature dependence of spontaneous magnetization at temperature approaching the ordering temperature from below let's expand all Brillouin functions at small $x$

$$B_S(x) = \frac{(2S+1)^2 - 1}{(2S)^2}\frac{x}{3} - \frac{(2S+1)^4 - 1}{3(2S)^4}\frac{x^3}{45}, \qquad x \ll 1. \tag{22}$$

The system of equations for sub-lattices magnetizations (18) is essentially simplified if one chooses the concrete type of ordering.

1. B - structure.

As it can be seen from Fig.2, at ferromagnetic ordering one has

$$x_3 = x_5 = x_7 = x_1,$$

$$x_4 = x_6 = x_8 = x_2 \quad . \tag{23}$$

As a result system (18) will consist of two equations

$$\begin{aligned} x_1 &= S_1 B_{S_1}(Z_{11}) \\ x_2 &= S_2 B_{S_2}(Z_{21}), \end{aligned} \tag{24}$$

where

$$\begin{aligned} Z_{11} &= \alpha x_1 + \beta x_2, \\ Z_{21} &= \gamma x_1 + \delta x_2, \end{aligned} \tag{25}$$

$$\alpha = \frac{2S_1}{T}(1-x)^2(4I_1 + 2I_2), \tag{26}$$

$$\beta = \frac{2S_1}{T}x(1-x)(4\tilde{I}_5 + 2\tilde{I}_6), \tag{27}$$

$$\gamma = \frac{2S_2}{T}x(1-x)(4\tilde{I}_5 + 2\tilde{I}_6), \tag{28}$$

$$\delta = \frac{2S_2}{T}x^2(4I_3 + 2I_4). \tag{29}$$

As a result we obtain the following system for magnetizations of the first sub-lattice $x_1$ and $x_2$ near Curie temperature

$$x_1 = t_1(\alpha x_1 + \beta x_2) - a_1(\alpha x_1 + \beta x_2)^3$$
$$x_2 = t_2(\gamma x_1 + \delta x_2) - a_2(\gamma x_1 + \delta x_2)^3, \quad (30)$$

where

$$t_1 = \frac{S_1 + 1}{3}, \quad (31)$$

$$t_2 = \frac{S_2 + 1}{3}, \quad (32)$$

$$a_1 = \frac{(S_1 + 1)[1 + (2S_1 + 1)^2]}{45(2S_1)^2}, \quad (33)$$

$$a_2 = \frac{(S_2 + 1)[1 + (2S_2 + 1)^2]}{45(2S_2)^2}. \quad (34)$$

The iteration method gives the following solution of system (30)

$$x_1{}^2 = \frac{1}{\gamma^2} \frac{|b| - \sqrt{b^2 - 4ac}}{2a}, \quad (35)$$

where

$$a = 3a_1 a_2 t_2{}^2 \beta^2, \quad (36)$$

$$b = -[a_1 t_2^3 \beta^2 + a_2 t_1 (1 - t_2\delta)^2] \ , \quad (37)$$

$$c = \frac{(1 - t_2\delta)^2}{\beta\gamma} [t_1 t_2 \beta\gamma - (1 - t_1\alpha)(1 - t_2\beta)]. \quad (38)$$

When temperature approaches the ordering temperature from below magnetization $x_1$ should turn to zero. One can see that this is carried out when the condition $c = 0$ is fulfilled. Solution of this equation leads to the following expression for Curie temperature

$$T_C = \frac{1}{3}\left\{M_B + \sqrt{N_B + P_B}\right\} , \quad (39)$$

where

$$M_B = \left[(1-x)^2 S_1(S_1 + 1)(4I_1 + 2I_2) + x^2 S_2(S_2 + 1)(4I_3 + 2I_4)\right] \ , \quad (40)$$

$$N_B = \left[(1-x)^2 S_1(S_1+1)(4I_1+2I_2) - x^2 S_2(S_2+1)(4I_3+2I_4)\right]^2 \quad , \tag{41}$$

$$P_B = 4x^2(1-x)^2 S_1(S_1+1)S_2(S_2+1)(4\tilde{I}_5+2\tilde{I}_6)^2. \tag{42}$$

One can easy see that magnetization tends to zero at approaching $T_c$ according to the law

$$x_1^2 \approx \frac{1}{\gamma^2(T_c)b(T_c)}\frac{\partial c}{\partial T}\bigg|_{T_c}(T-T_c). \tag{43}$$

2. A - structure.

At antiferromagnetic ordering of A – type we have

$$\begin{aligned} &x_3 = x_1, x_5 = x_7 = -x_1, \\ &x_4 = x_2, x_6 = x_8 = -x_2. \end{aligned} \tag{44}$$

The iteration method analogous to that for ferromagnetic B-ordering leads to the following expression for Neel temperature

$$T_N^A = \frac{1}{3}\left\{M_A + \sqrt{N_A + P_A}\right\}, \tag{45}$$

where

$$M_A = \left[(1-x)^2 S_1(S_1+1)(4I_1-2I_2) + x^2 S_2(S_2+1)(4I_3-2I_4)\right], \tag{46}$$

$$N_A = \left[(1-x)^2 S_1(S_1+1)(4I_1-2I_2) - x^2 S_2(S_2+1)(4I_3-2I_4)\right]^2 , \tag{47}$$

$$P_A = 4x^2(1-x)^2 S_1(S_1+1)S_2(S_2+1)(4\tilde{I}_5-2\tilde{I}_6)^2. \tag{48}$$

3. C -structure.

As it can be seen from Fig.2, the following correlations between sub-lattices magnetizations take place in this case

$$\begin{aligned} &x_3 = -x_1, x_5 = x_1, x_7 = -x_1, \\ &x_4 = -x_2, x_6 = x_2, x_8 = -x_2. \end{aligned} \tag{49}$$

Then the Neel temperature has the form

$$T_N^C = \frac{1}{3}\left\{M_C + \sqrt{N_C + P_C}\right\}, \tag{50}$$

where

$$M_C = \left[(1-x)^2 S_1(S_1+1)(-4I_1+2I_2) + x^2 S_2(S_2+1)(-4I_3+2I_4)\right] , \tag{51}$$

$$N_C = \left[(1-x)^2 S_1(S_1+1)(4I_1-2I_2) - x^2 S_2(S_2+1)(4I_3-2I_4)\right]^2 , \tag{52}$$

$$P_C = 4x^2(1-x)^2 S_1(S_1+1)S_2(S_2+1)(4\tilde{I}_5 - 2\tilde{I}_6)^2. \tag{53}$$

3. G - structure.

With the use of conditions

$$\begin{aligned} x_3 &= -x_1, x_5 = -x_1, x_7 = x_1, \\ x_4 &= -x_2, x_6 = -x_2, x_8 = x_2 \end{aligned} \tag{54}$$

we obtain the following expression for ordering temperature

$$T_N^G = \frac{1}{3}\left\{M_G + \sqrt{N_G + P_G}\right\}, \tag{55}$$

where

$$M_G = \left[-(1-x)^2 S_1(S_1+1)(4I_1+2I_2) - x^2 S_2(S_2+1)(4I_3+2I_4)\right], \tag{56}$$

$$N_G = \left[(1-x)^2 S_1(S_1+1)(4I_1+2I_2) - x^2 S_2(S_2+1)(4I_3+2I_4)\right]^2, \tag{57}$$

$$P_G = 4x^2(1-x)^2 S_1(S_1+1)S_2(S_2+1)(4\tilde{I}_5 + 2\tilde{I}_6)^2. \tag{58}$$

For all three collinear antiferromagnetic structures A, C and G the temperature dependence of separate sub-lattice magnetization near ordering temperature has the form

$$x_1 \approx \sqrt{T_N^i - T}, \qquad i = A, C, G. \tag{59}$$

# IV. TEMPERATURE DEPENDENCE OF SEPARATE SUB-LATTICE MAGNETIZATION IN THE WHOLE TEMPERATURE RANGE

When we investigated the temperature dependence of magnetization near ordering temperature where magnetization tends to zero the square root in the expression of double exchange operator (5) was expanded into series. If we are far from ordering temperature such an expansion is not righteous. The magnetization is expressed via Brillouin function only at Heisenberg form of the Hamiltonian. In our case magnetization should be calculated with the help of overall formula

$$\langle J_z \rangle = \frac{Sp(J_z \exp(-\hat{H}/T))}{Sp(\exp(-\hat{H}/T)}. \quad (60)$$

In the molecular field approximation the systems of equations for mean spin values of $Mn^{3+}$ and $Mn^{4+}$ ions, belonging to the first sub-lattice at different types of ordering have the following form

B-structure

$$x_1 = \frac{\sum_{m=-2}^{2} m \exp\left\{\frac{2(1-x)}{T}\left[K_B + L_B\right]\right\}}{\sum_{m=-2}^{2} \exp\left\{\frac{2(1-x)}{T}\left[K_B + L_B\right]\right\}},$$

$$(61)$$

$$x_2 = \frac{\sum_{n=-3/2}^{3/2} n \exp\left\{\frac{2x}{T}\left[R_B + Q_B\right]\right\}}{\sum_{m=-3/2}^{3/2} \exp\left\{\frac{2x}{T}\left[R_B + Q_B\right]\right\}},$$

$$K_B = m(1-x)(4I_1 + 2I_2)x_1 + x(4I_5 + 2I_6)x_2, \quad (62)$$

$$L_B = x\sqrt{\frac{S_2+1}{2S_2+1}}(4B_1 + 2B_2)\sqrt{1+\eta x_2 m}, \quad (63)$$

$$R_B = n(1-x)(4I_5 + 2I_6)x_1 + x(4I_3 + 2I_4)x_2, \quad (64)$$

$$Q_B = (1-x)\sqrt{\frac{S_2+1}{2S_2+1}}(4B_1 + 2B_2)\sqrt{1+\eta x_1 n}, \quad (65)$$

$$\eta = \frac{2}{1 + S_2(2S_2+3)}. \quad (66)$$

A-structure

$$x_1 = \frac{\sum_{m=-2}^{2} m \exp\left\{\frac{2(1-x)}{T}\left[K_A + L_A\right]\right\}}{\sum_{m=-2}^{2} \exp\left\{\frac{2(1-x)}{T}\left[K_A + L_A\right]\right\}},$$

$$(67)$$

$$x_2 = \frac{\sum_{n=-3/2}^{3/2} n \exp\left\{\frac{2x}{T}\left[R_A + Q_A\right]\right\}}{\sum_{m=-3/2}^{3/2} \exp\left\{\frac{2x}{T}\left[R_A + Q_A\right]\right\}},$$

$$K_A = m(1-x)(4I_1 - 2I_2)x_1 + x(4I_5 - 2I_6)x_2, \tag{68}$$

$$L_A = x\sqrt{\frac{S_2+1}{2S_2+1}}(4B_1\sqrt{1+\eta x_2 m} + 2B_2\sqrt{1-\eta x_2 m}), \tag{69}$$

$$R_A = n(1-x)(4I_5 - 2I_6)x_1 + x(4I_3 - 2I_4)x_2, \tag{70}$$

$$Q_A = (1-x)\sqrt{\frac{S_2+1}{2S_2+1}}(4B_1\sqrt{1+\eta x_1 n} + 2B_2\sqrt{1-\eta x_1 n}). \tag{71}$$

C-structure

$$x_1 = \frac{\sum_{m=-2}^{2} m\exp\left\{\frac{2(1-x)}{T}\left[K_C + L_C\right]\right\}}{\sum_{m=-2}^{2}\exp\left\{\frac{2(1-x)}{T}\left[K_C + L_C\right]\right\}},$$

$$x_2 = \frac{\sum_{n=-3/2}^{3/2} n\exp\left\{\frac{2x}{T}\left[R_C + Q_C\right]\right\}}{\sum_{m=-3/2}^{3/2}\exp\left\{\frac{2x}{T}\left[R_C + Q_C\right]\right\}}. \tag{72}$$

$$K_C = -m(1-x)(4I_1 - 2I_2)x_1 + x(4I_5 - 2I_6)x_2, \tag{73}$$

$$L_C = x\sqrt{\frac{S_2+1}{2S_2+1}}(4B_1\sqrt{1-\eta x_2 m} + 2B_2\sqrt{1+\eta x_2 m}), \tag{74}$$

$$R_C = -n(1-x)(4I_5 - 2I_6)x_1 + x(4I_3 - 2I_4)x_2, \tag{75}$$

$$Q_C = (1-x)\sqrt{\frac{S_2+1}{2S_2+1}}(4B_1\sqrt{1-\eta x_1 n} + 2B_2\sqrt{1+\eta x_1 n}). \tag{76}$$

G-structure

$$x_1 = \frac{\sum_{m=-2}^{2} m\exp\left\{\frac{2(1-x)}{T}\left[K_G + L_G\right]\right\}}{\sum_{m=-2}^{2}\exp\left\{\frac{2(1-x)}{T}\left[K_G + L_G\right]\right\}},$$

$$x_2 = \frac{\sum_{n=-3/2}^{3/2} n\exp\left\{\frac{2x}{T}\left[R_G + Q_G\right]\right\}}{\sum_{m=-3/2}^{3/2}\exp\left\{\frac{2x}{T}\left[R_G + Q_G\right]\right\}}, \tag{77}$$

$$K_G = -m(1-x)(4I_1 + 2I_2)x_1 + x(4I_5 + 2I_6)x_2, \quad (78)$$

$$L_G = x\sqrt{\frac{S_2+1}{2S_2+1}}(4B_1 + 2B_2)\sqrt{1-\eta x_2 m}), \quad (79)$$

$$R_G = -n(1-x)(4I_5 + 2I_6)x_1 + x(4I_3 + 2I_4)x_2, \quad (80)$$

$$Q_G = (1-x)\sqrt{\frac{S_2+1}{2S_2+1}}(4B_1 + 2B_2)\sqrt{1-\eta x_1 n}). \quad (81)$$

The systems of equations (61), (67), (72) and (77) were solved numerically with the use of exchange parameters and transfer integrals values found in paper [1]: $I_1 = 9.6K$, $I_2 = -6.7K$, $I_3 = I_4 = -8.7K$, $I_5 = -12.7K$, $I_6 = 9.6K$, $B_1 = 144.6K$, $B_2 = 116.1K$. At that for each type of ordering B, A, C or G some concentration was chosen from the corresponding interval. Figs.3-6 display the temperature dependences of mean values $x_1$ and $x_2$ of ions $Mn^{3+}$ and $Mn^{4+}$ spins, correspondingly. Figs.7-10 describe temperature dependences of contributions to magnetization on separate site from ions $Mn^{3+}$

$$\sigma_1 = (1-x)x_1 \quad (82)$$

and $Mn^{4+}$

$$\sigma_2 = xx_2. \quad (83)$$

In the same place on Figs.7-10 temperature dependence of the total magnetization of a separate site

$$\sigma = \sigma_1 + \sigma_2 \quad (84)$$

is presented for all investigated collinear configurations. It should be noted that for configurations B and C contributions to separate site magnetization from ions $Mn^{3+}$ and $Mn^{4+}$ have one and the same sign while for configurations A and G the signs are different.

## RESUME

The magnetic properties of four collinear magnetic configurations of compound $La_{1-x}Ca_xMnO_3$ are investigated in molecular field approximation. Expressions for ordering temperatures are found. They have the form analogous to Neel tempera-ture for ferrimagnets [3], and contain evident dependence upon concentration $x$ of $Mn^{4+}$ ions.

The analytical temperature dependences of spontaneous site magnetization were obtained at temperatures near the ordering temperature. For all collinear structures under investigation this dependence has the form $\sigma \approx \sqrt{T_{ord} - T}$ .

Temperature dependences of site magnetization are calculated in the whole temperature range from zero till ordering temperature.

## ACKNOWLEDGEMENTS

This work was fulfilled in the framework of Department of Physical Science of Russian Academy of Sciences and the Program of Presidium of Ural Branch of Russian Academy of Sciences “New materials and structures”.

## FIGURE CAPTIONS

Fig.1. The elementary cell of $La_{1-x}Ca_xMnO_3$. Only Mn-sites are denoted. Figures 1-4 number Bravais lattices.

Fig.2. Collinear magnetic configurations $A, B, C, G$. Arrows denote the site spins.

Fig.3. Temperature dependence of mean spin values for ions $Mn^{3+}$ (short dash-curve) and $Mn^{4+}$ (dash dot curve) at ferromagnetic ordering. The data are obtained with concentration $x = 0.3$, in this case $T_c = 55.9K$.

Fig.4. Temperature dependence of mean spin values for ions $Mn^{3+}$ (short dash-curve) and $Mn^{4+}$ (dash dot curve) at antiferromagnetic ordering of A-type ($x = 0.1, T_N^A = 168.7K$).

Fig.5. Temperature dependence of mean spin values for ions $Mn^{3+}$ (short dash-curve) and $Mn^{4+}$ (dash dot curve) at antiferromagnetic ordering of C-type ($x = 0.75, T_N^C = 37.2K$).

Fig.6. Temperature dependence of mean spin values for ions $Mn^{3+}$ (short dash-curve) and $Mn^{4+}$ (dash dot curve) at antiferromagnetic ordering of G-type ($x = 0.9, T_N^A = 106.5K$).

Fig.7. Temperature dependence of contributions to magnetization of a separate site from ions $Mn^{3+}$ (short dash curve) and $Mn^{4+}$ (dash dot curve) and the total magnetization of a separate site (solid curve) at ferromagnetic ordering.

Fig.8. Temperature dependence of contributions to magnetization of a separate site from ions $Mn^{3+}$ (short dash curve) and $Mn^{4+}$ (dash dot curve) and the total magnetization of a separate site (solid curve) at antiferromagnetic ordering of A-type.

Fig.9. Temperature dependence of contributions to magnetization of a separate site from ions $Mn^{3+}$ (short dash curve) and $Mn^{4+}$ (dash dot curve) and the total magnetization of a separate site (solid curve) at antiferromagnetic ordering of C-type.

Fig.10. Temperature dependence of contributions to magnetization of a separate site from ions $Mn^{3+}$ (short dash curve) and $Mn^{4+}$ (dash dot curve) and the total magnetization of a separate site (solid curve) at antiferromagnetic ordering of G-type.

**Fig. 1**

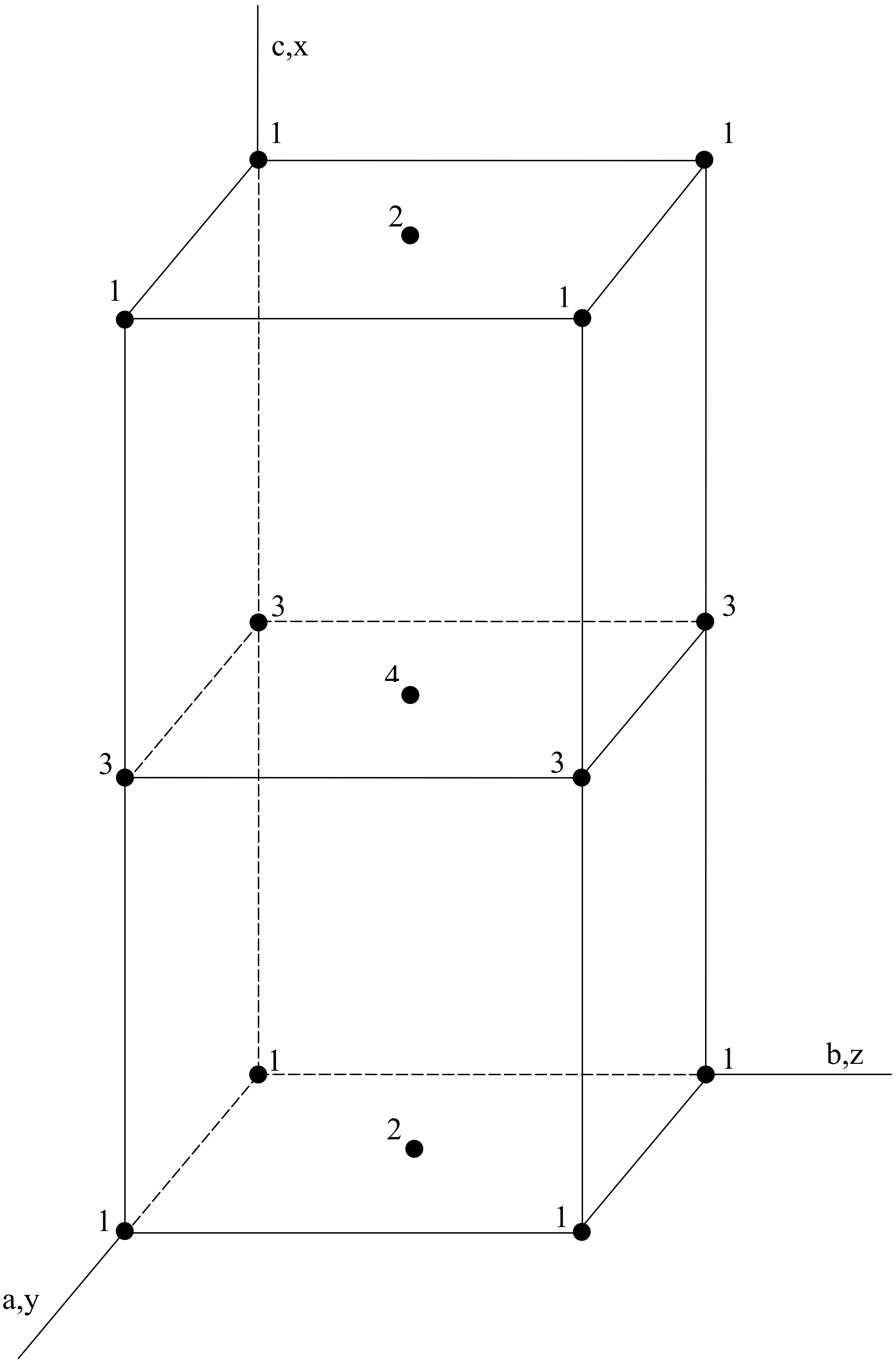

**Fig. 2**

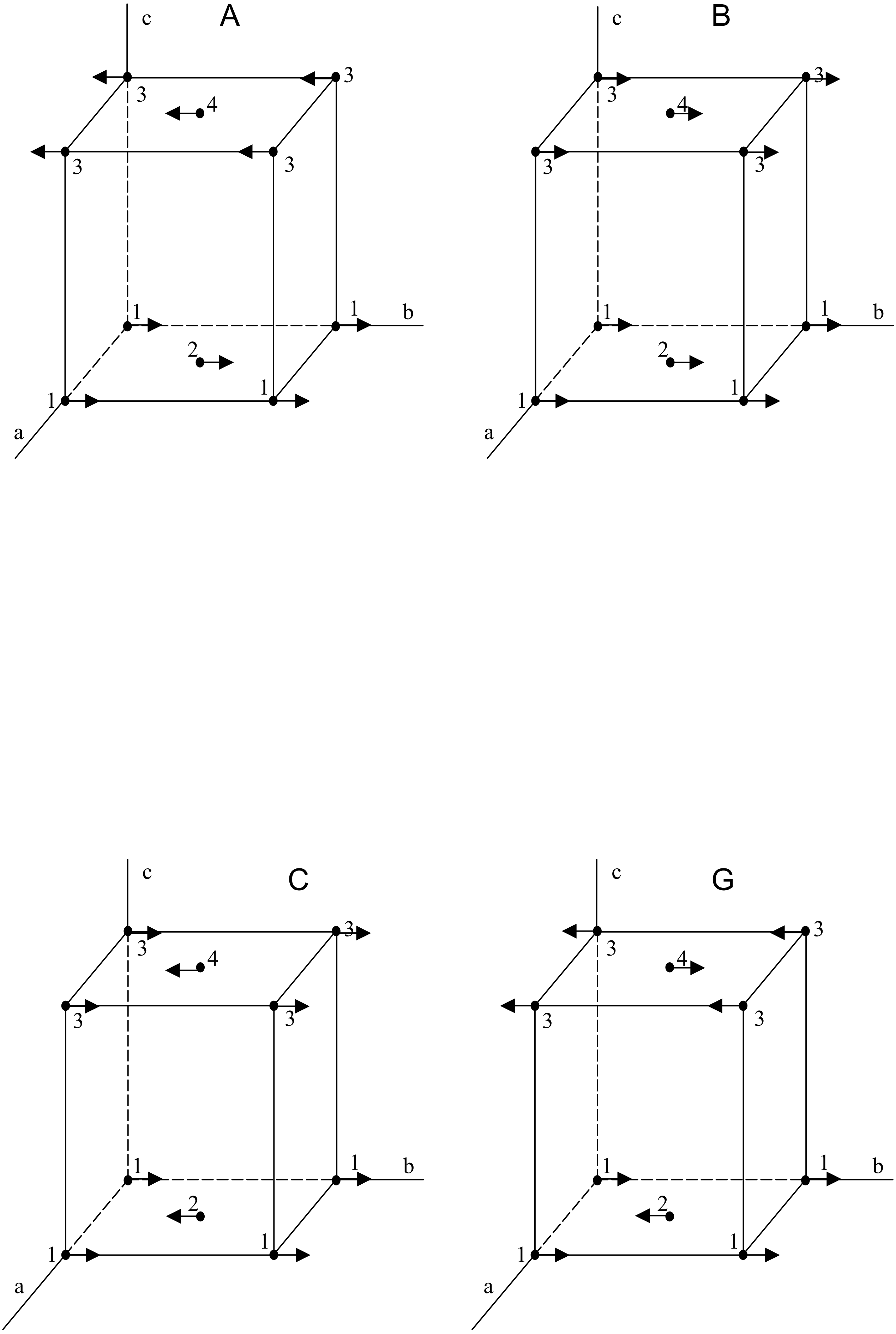

**Fig. 3**

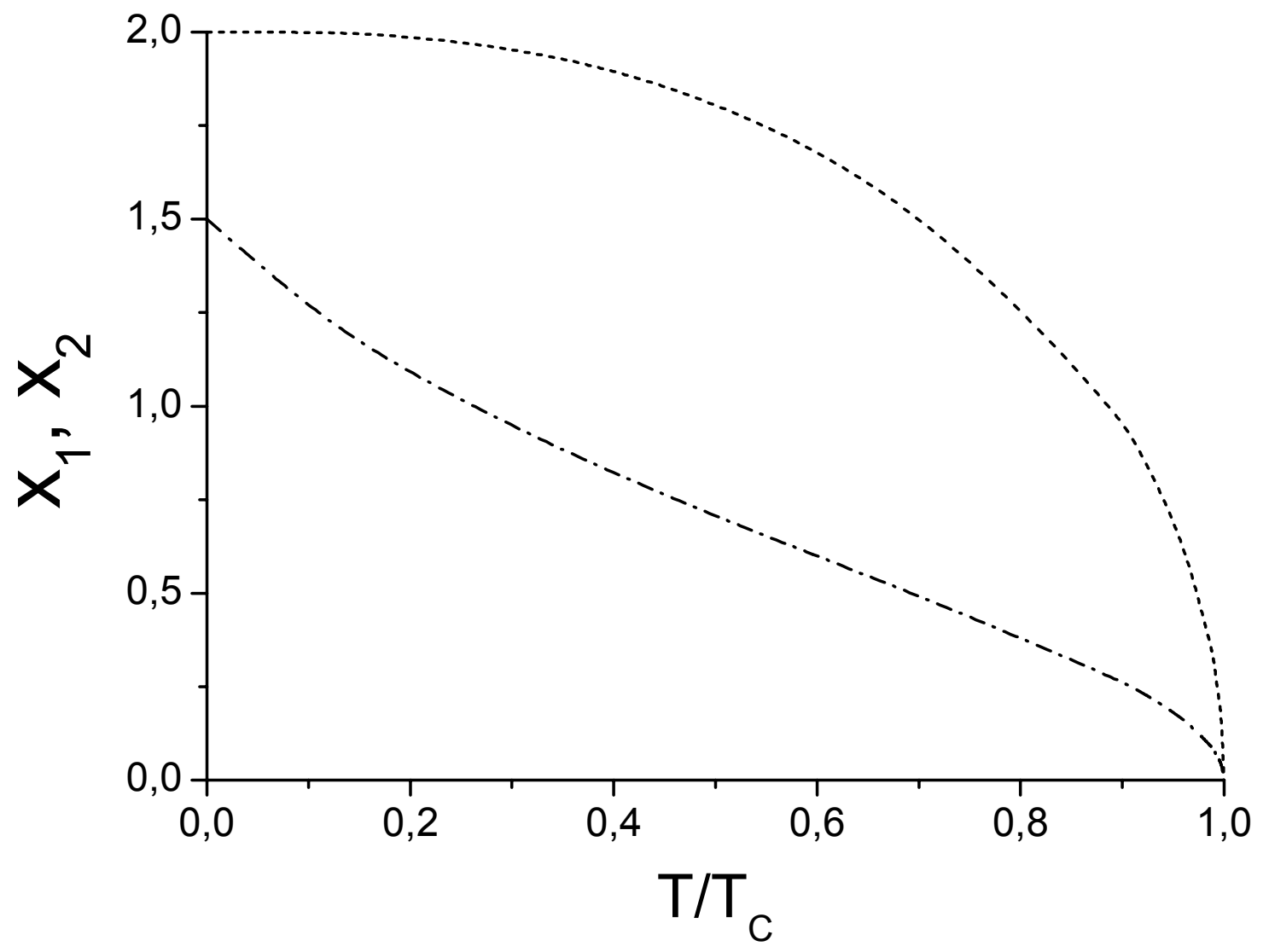

**Fig. 4**

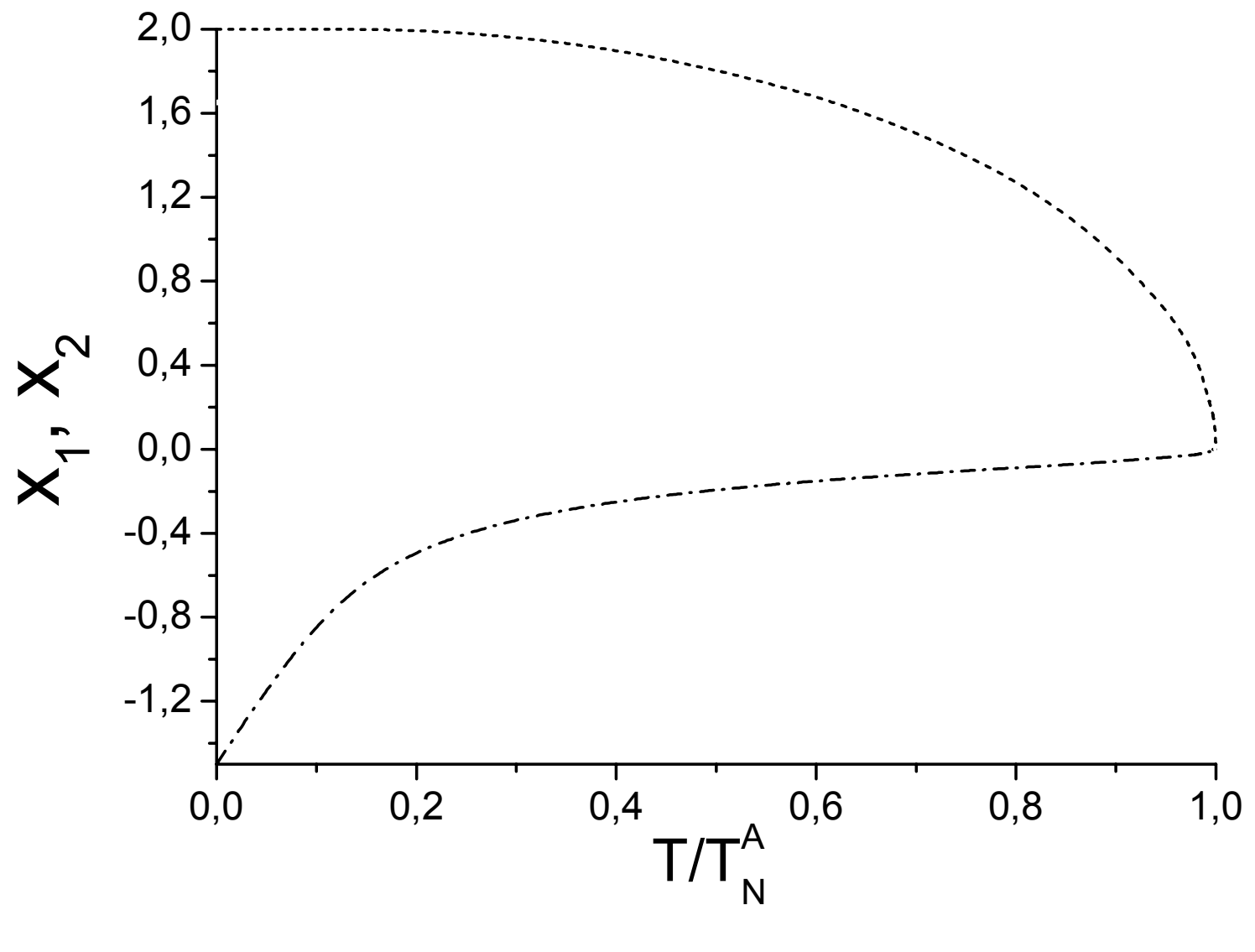

**Fig. 5**

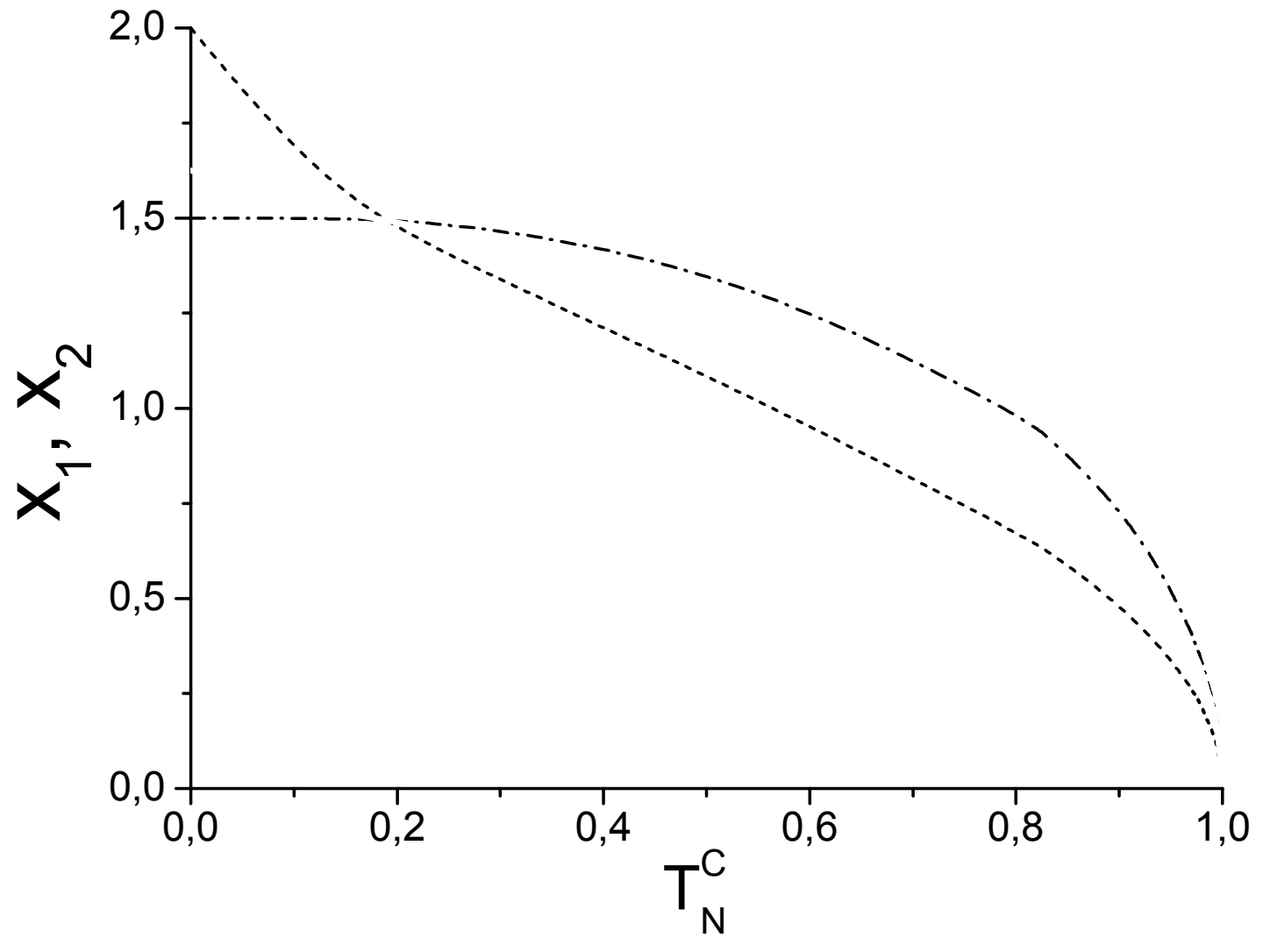

**Fig. 6**

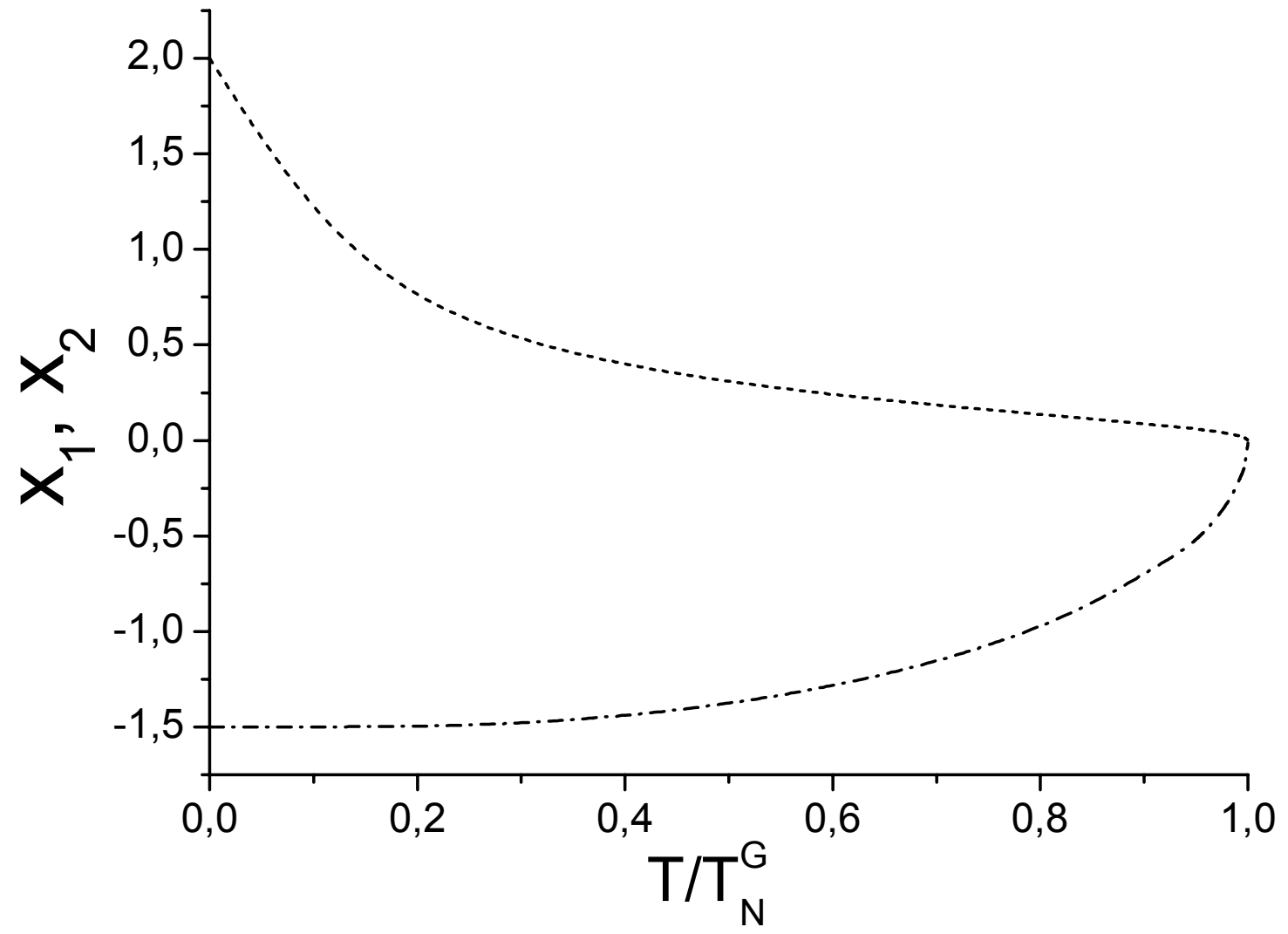

.

**Fig.7**

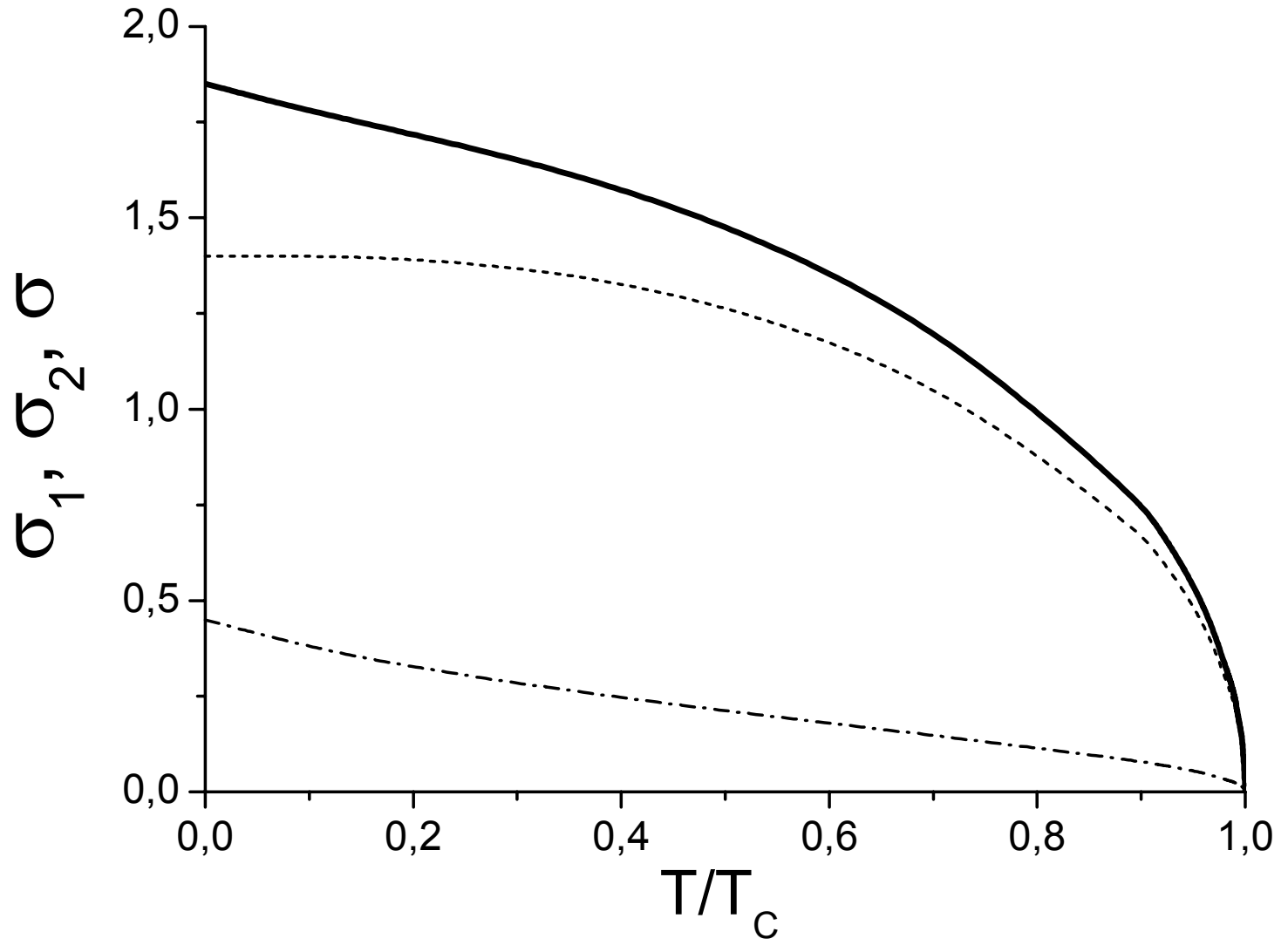

**Fig.8**

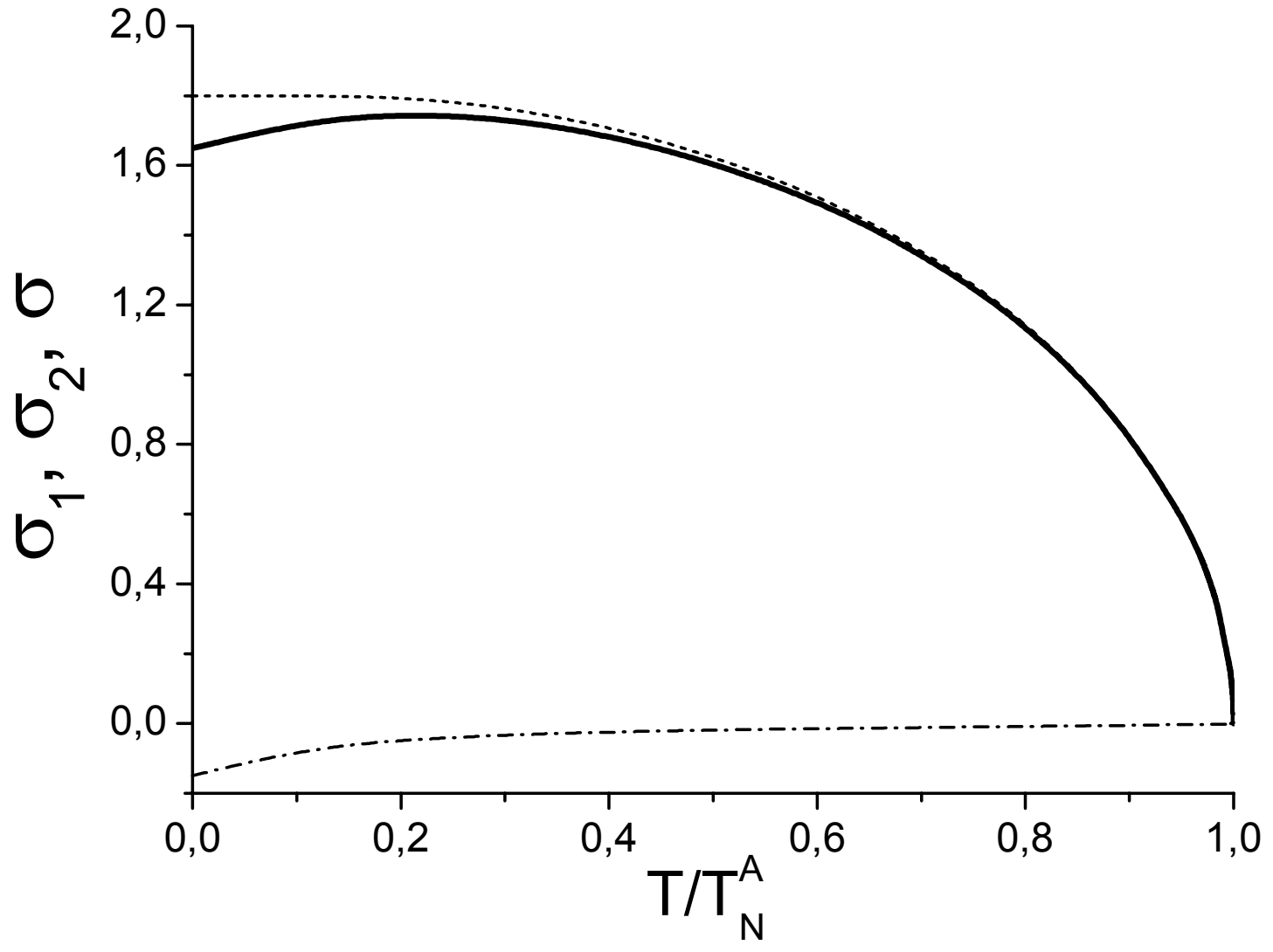

**Fig.9**

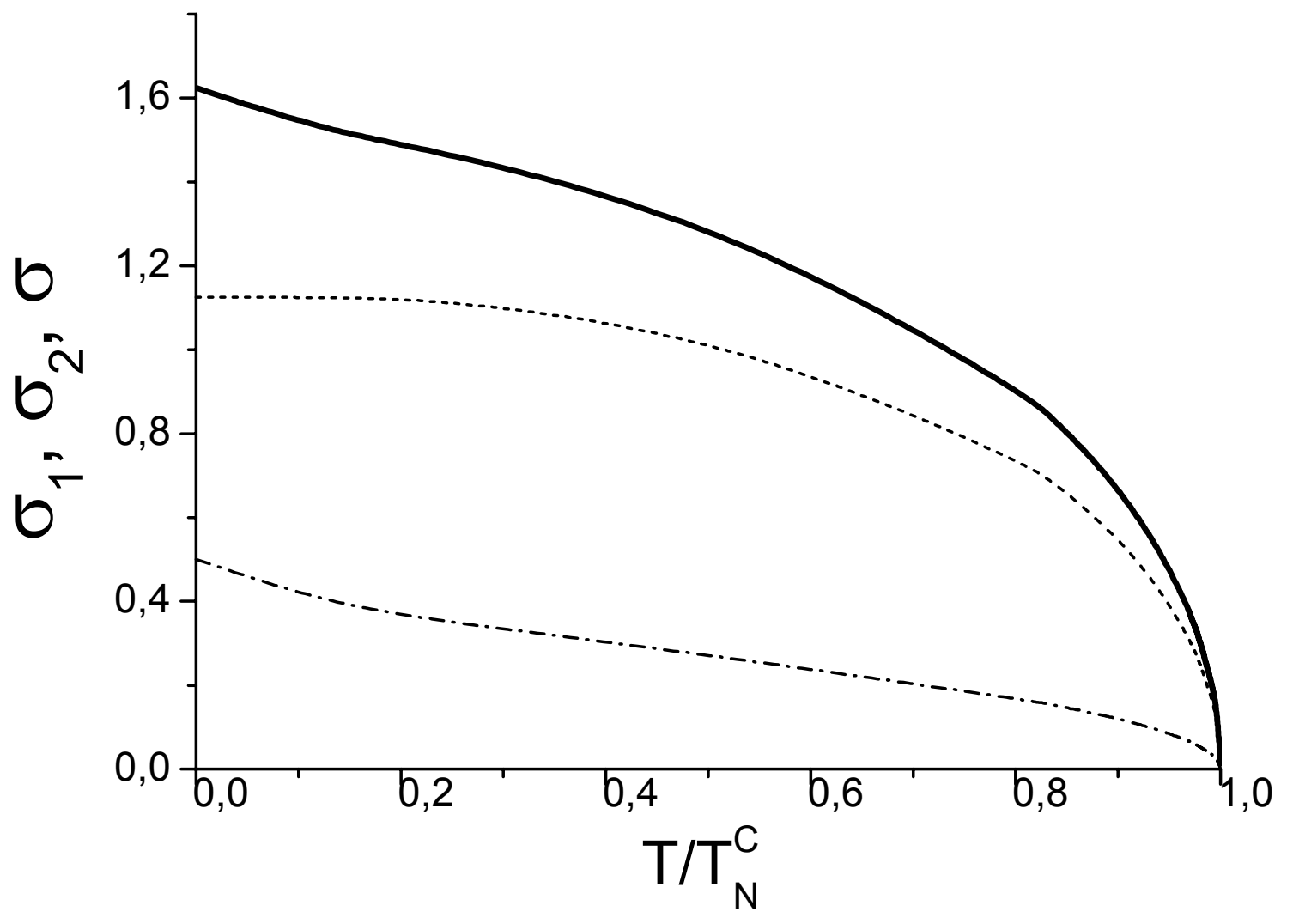

**Fig.10**

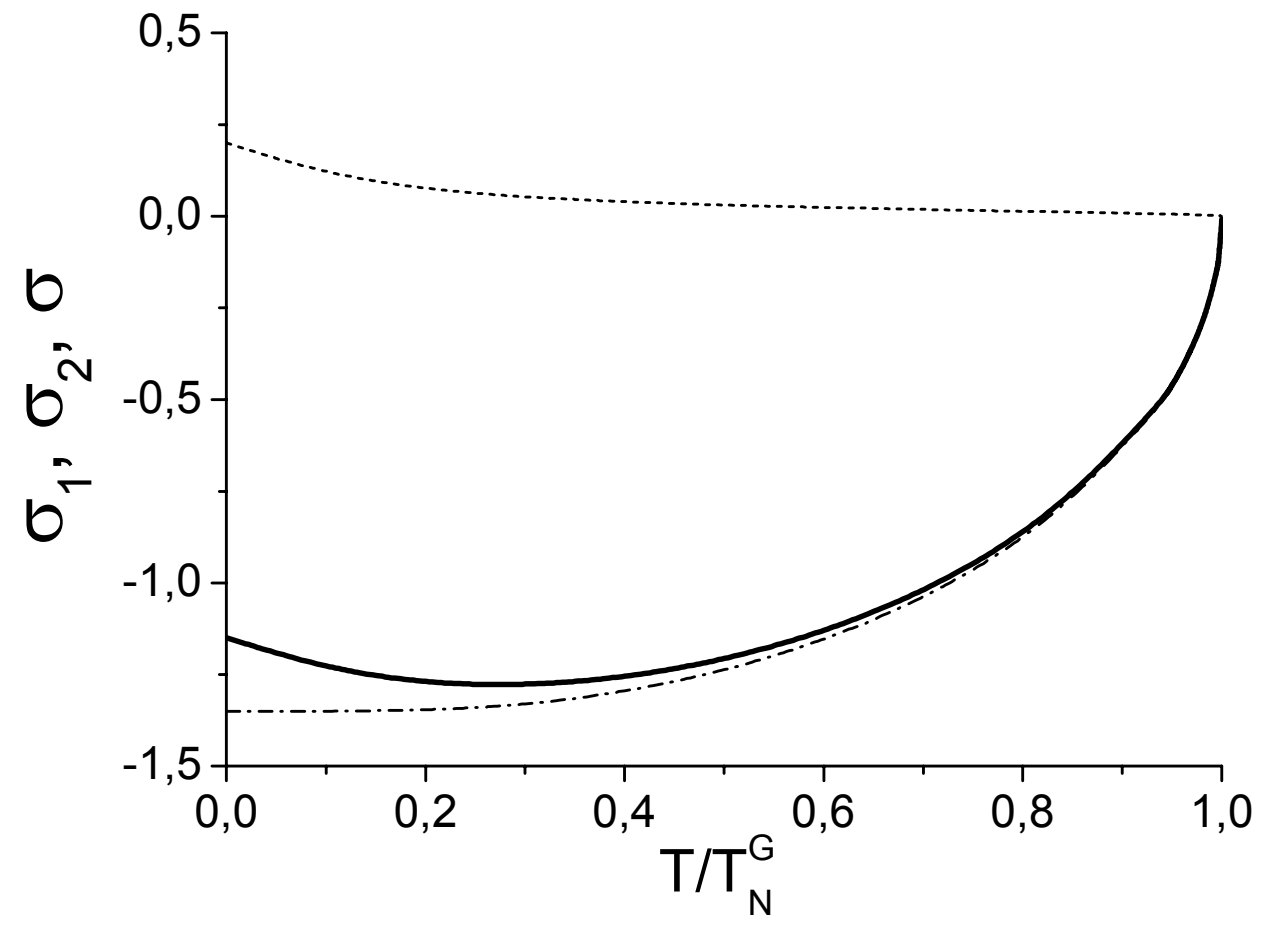